\documentclass[prb,showpacs,twocolumn,aps,superscriptaddress,a4paper]{revtex4-1}
\usepackage{dcolumn,amssymb,amsmath,amsfonts,graphicx,latexsym,color}

\begin{document}

\title{Identify the topological superconducting order in a multi-band quantum wire}
\author{Pei Wang*}
\email{wangpei@zjut.edu.cn}
\affiliation{International Center for Quantum Materials, School of Physics, Peking University, Beijing 100871, China}
\affiliation{Institute of Applied Physics, Zhejiang University of Technology, Hangzhou 310023, China}

\author{Jie Liu*}
\affiliation{Applied Physics Department, Xi'an Jiaotong University, Xi'an 710049, china}

\author{Qing-feng Sun}
\affiliation{International Center for Quantum Materials, School of Physics, Peking University, Beijing 100871, China}
\affiliation{Collaborative Innovation Center of Quantum Matter, Beijing 100871, China}

\author{X. C. Xie}
\affiliation{International Center for Quantum Materials, School of Physics, Peking University, Beijing 100871, China}
\affiliation{Collaborative Innovation Center of Quantum Matter, Beijing 100871, China}

\date{\today}

\begin{abstract}
How to distinguish the zero-bias peak (ZBP) caused by the Majorana fermions from that by
the other effects remains a challenge in detecting the topological order of a quantum wire.
In this paper we propose to distinguish the topological superconducting phase from the 
topologically trivial phase by making a Josephson junction of the quantum wire
attached to a side lead and then measuring the tunneling conductance through it
as the phase difference across the junction $\phi$ varies.
Even if the ZBPs exist in both phases, we can identify the topological superconducting
phase by a conductance peak at $\phi=\pi$ and a nearby butterfly pattern.
\end{abstract}

\maketitle

\emph{Introduction}.--
Recently the topological superconductor (TSC) has been a focus of attention
because it hosts the Majorana fermions which have potential applications in the
topological quantum computation. Among various proposals of realizing the TSC, a quantum wire
with strong spin-orbit coupling and proximity-induced superconductivity~\cite{sau10,alicea10,lutchyn10,oreg10}
attracts particular interests due to its easily being implemented.
While the nanowires of InSb and InAs~\cite{mourik,deng,das,churchill13}
have been manufactured as the candidates of TSCs,
the detection of topological order and Majorana fermions
keeps a highly nontrivial problem.

A TSC quantum wire has the Majorana fermions as its edge excitations.
The Majorana fermion leads to a zero-bias peak (ZBP) in the tunneling
conductance~\cite{law09,flensberg10,wimmer11,sau10b,akhmerov11},
which was recently observed in experiments~\cite{mourik,deng,das,churchill13}.
However, the ZBP by itself cannot exclusively sign a topologically nontrivial
Majorana fermion, since the disorder in quantum wires
may induce the sub-gap Andreev levels close to zero energy, which cause similar
ZBPs in the topologically trivial phase~\cite{bagrets12,liu12}.
Furthermore, the observed ZBP may also be caused by the Kondo effect~\cite{lee12}, 0.7 anomaly~\cite{churchill13}
or weak antilocalization~\cite{pikulin12}.
Only the Majorana fermions in the TSC phase can be used for the
quantum computation~\cite{alicea11,romito12,sau11,heck12},
it is then obliged to study how to distinguish the topologically trivial and nontrivial phases in the quantum wire.
Some authors invoked the non-local properties of Majorana fermions exhibited
in the current-current correlations~\cite{bolech07,bose,liu13}
or the dependence of the ZBP on the confining potential at the opposite end of the wire~\cite{stanescu}.
The others kept on exploiting the local properties of Majorana fermions in
the selective equal-spin Andreev reflections~\cite{he14},
the conductance of a quantum dot that is coupled
to a TSC~\cite{liude}, the scaling property of the conductance at finite temperatures
with a resistive lead~\cite{liuDE13},
or the tunneling process with the ferromagnetic leads~\cite{ren}.
Whereas these suggestions either require a measurement of the correlations which is difficult to implement
or do not systematically consider the effect of
the casual Andreev bound level whose local properties are similar to those of a Majorana fermion.


\begin{figure}
\includegraphics[width=0.9\linewidth]{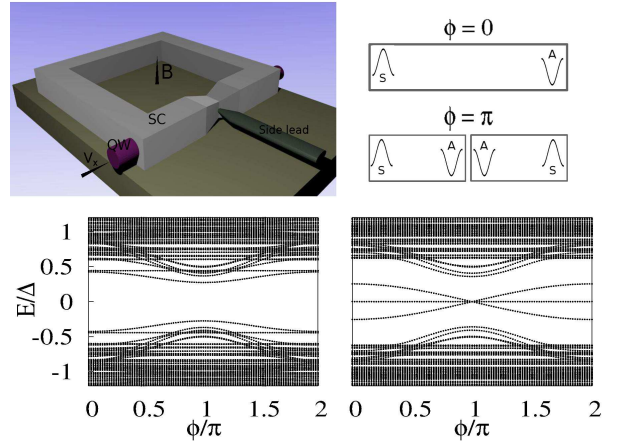}
\caption{[Color online]. [Left top panel] The schematic diagram of our proposal. A quantum wire (QW)
is covered by the ring-shaped superconducting layer (SC) which forms a Josephson junction
at the middle of the wire. A side lead is connected to the junction to measure the tunneling conductance.
A parallel magnetic field $V_x$ is necessary for generating the topological order.
While a transverse field $B$ is applied to tune the phase difference $\phi$
across the junction. [Right top panel] There are
a symmetric Majorana fermion ($S$) and an antisymmetric one ($A$) at the ends of the wire,
respectively, if the superconducting phase is uniform throughout the wire.
But if a $\pi$-phase difference is across the junction, two additional Majorana fermions with the same symmetry
will emerge at the junction. [Bottom panels] The eigenenergy spectra of $\hat H_w$
in the TSC phase at $\mu=-2t$ (right bottom panel)
and in the trivial phase at $\mu=-2t+4\Delta$ (left bottom panel).}\label{fig:schematic}
\end{figure}

In this paper, we propose to distinguish the topologically trivial and nontrivial phases
in a quantum wire by coupling a
side lead (or a STM tip) to its center where the proximity-induced superconductivity is weakened
to a weak coupling, i.e., a Josephson junction.
Fig.~\ref{fig:schematic} (left top panel) shows the schematic diagram of our proposal.
It was known~\cite{kitaev,kwon04,fu09,jiang11,lutchyn10} that in the TSC phase a pair of
Majorana fermions emerge at the center of the Josephson junction with the coupling energy
proportional to $cos(\phi/2)$ where $\phi$ is the phase difference across the junction.
These Majorana fermions induce the well-known fractional Josephson effect (FJE).
Several experiments were conducted for observing the putative
$4\pi$-periodic FJE~\cite{yacob, kou}, but failed due to the parity-flipping processes.
It is surprising that we can distinguish these Majorana fermions
from the trivial zero-energy Andreev bound states with the help of a side lead,
because the Majorana fermions have the same symmetry in their particle-hole 
wave functions (Fig.~\ref{fig:schematic}, right top panel)
which leads to a constructive interference in the tunneling process.
Even if the ZBP exists in both the TSC and the trivial phases, it
splits into three peaks in the TSC phase as $\phi$ deviates from $\pi$, displaying
a butterfly pattern, but keeps invariant in the trivial phase.
A conductance peak at $\phi=\pi$ and a nearby butterfly pattern are the fingerprints of the TSC phase.

The key difference between our proposal and previous ones is
the utilization of an additional degree of freedom, i.e., the phase difference $\phi$, which can be
tuned by applying a transverse magnetic field $\textbf{B}$.
This magnetic field is screened by
the superconducting layer and does not affect the quantum wire directly. The effects
disguising themselves as Majorana fermions, e.g., the disorder effect, the Kondo effect, etc.,
are indifferent to the superconducting phase, so that they can be ruled out by the conductance peak at $\phi=\pi$.
Finally, our proposal only requires measuring the tunneling conductance through the side-lead
and is practicable in current techniques.

\emph{Theoretical model of the quantum wire}.--
We simulate the quantum wire in the $x$ direction by a rectangular lattice of size
$\left(N_x\times N_y \right)$ lying in the $x-y$ plane with the Hamiltonian:
\begin{equation}
\begin{split}
 \hat H_{w} = & \sum_{\vec{R},\vec{d},\alpha} -t (\vec{R},\vec{d}) \hat \psi^\dag_{\vec{R}+\vec{d}, \alpha}
 \hat \psi_{\vec{R},\alpha} - \mu \sum_{\vec{R},\alpha} \hat \psi^\dag_{\vec{R},\alpha}
 \hat \psi_{\vec{R},\alpha}
 \\ & -i U_R \sum_{\vec{R},\vec{d},\alpha,\beta}  \hat \psi^\dag_{\vec{R}+\vec{d}, \alpha}
 \left( \vec{z} \cdot \left( \vec{ \sigma }_{\alpha\beta} \times \vec{d} \right) \right) \hat \psi_{\vec{R},\beta}
 \\ & + V_x \sum_{\vec{R},\alpha,\beta} \hat \psi^\dag_{\vec{R},\alpha} \left(\sigma_x\right)_{\alpha\beta}
 \hat \psi_{\vec{R},\beta} \\
 & + \sum_{\vec{R},\alpha}
 \left( \Delta(\vec{R}) \hat \psi^\dag_{\vec{R},\alpha} \hat \psi^\dag_{\vec{R},-\alpha} + h.c.\right) ,
\end{split}
\end{equation}
where $\vec{R}$ denotes the lattice sites, $\vec{d}$ the unit vectors connecting the nearest neighbor sites
in the $x$ and $y$ directions, $\alpha$ and $\beta$ the spins, $\mu$ the chemical potential,
$U_R$ the Rashba coupling, $\vec{z}$ the unit vector along the $z$ direction, and $\vec{\sigma}$
the Pauli matrices. $V_x$ is a magnetic field along the
wire for generating the topological order.
The on-site superconducting pairing $\Delta(\vec{R})$
equals to $\Delta$ and $\Delta e^{i\phi}$ in the left and right sides of the Josephson
junction respectively, with $\phi$ denoting
the phase difference across the junction which is tuned by a transverse magnetic field.
Finally, a barrier arises from the weakened superconductivity
at the junction, which is
simulated by a suppressed nearest-neighbor hopping, i.e.,
we set $t(\vec{R},\vec{d})=t$ for most $(\vec{R},\vec{d})$ but $t(\vec{R},\vec{d})=t'<t$ if $\vec{R}$
and $(\vec{R}+\vec{d})$ locate at different sides of the junction.
We emphasize that the weak coupling ($t'<t$) is a necessary condition for
distinguishing the TSC phase from the trivial phase by inducing an avoided crossing in the spectrum of the latter.
We choose the parameters in the tight-binding model that match the corresponding values in recent
experiments~\cite{liu12}: $\Delta=250 \mu \text{e}V$, $t=10\Delta$, $V_x=2\Delta$ and $U_{R}=2\Delta$,
and set the coupling at the junction to $t'=0.5 t$.
The dimensions of the wire are $N_x a = 200a \approx 4\mu m$ and $ N_y a = 5a \approx 100nm$.

\begin{figure}
\includegraphics[width=1.0\linewidth]{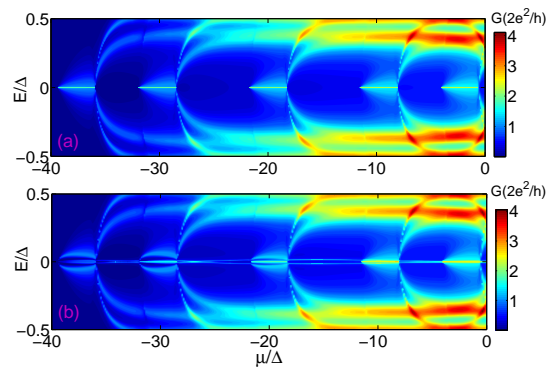}
\caption{[Color online] The 3D plot of the conductance $G$ as a function of the chemical potential $\mu$
and the voltage bias $E$ at (a) $\lambda=0,\epsilon=0.001t$ and (b) $\lambda=0.3, \epsilon=0.001t$.
The unit of $G$ is $2e^2/h$.}\label{fig:mu}
\end{figure}

The quasi-one-dimensional wire has $10$ transverse subbands.
It is a TSC so long as
an odd number of subbands are occupied~\cite{potter10,potter11}.
By varying the chemical potential $\mu$,
we observe a sequence of TSC phases well separated by the topologically trivial phases.
In the TSC phase, a pair of Majorana bound states are located at the ends of the wire
regardless of the value of $\phi$ (Fig.~\ref{fig:schematic}, right bottom panel).
At the same time, an additional pair of sub-gap Andreev levels
cross each other at $\phi=\pi$, corresponding to two
Majorana bound states located at the junction.
They have the same symmetry in their particle-hole wave functions, so that the coupling
between them vanishes at $\phi=\pi$.
In the topologically trivial phase, however, the sub-gap levels do not cross,
but avoid each other instead, so that there is no
Majorana bound states (Fig.~\ref{fig:schematic}, left bottom panel).

\emph{Topologically trivial sub-gap Andreev levels}.--
The disorder in a quantum wire may casually generate a pair of
Andreev bound levels near zero energy in the topologically trivial phase,
which cause a ZBP in the tunneling spectroscopy just like the Majorana fermions~\cite{liu12,bagrets12}.
For simulating these topologically trivial sub-gap Andreev levels,
we introduce an additional localized impurity state
\begin{equation}\label{manualAndreev}
\begin{split}
 \hat H_{d} = \epsilon \hat d^\dag \hat d + \lambda( \hat d^\dag \hat \psi_{\vec{R}_J, \alpha} + h.c.),
\end{split}
\end{equation}
where $\hat d$ denotes the field operator of the impurity state located at $\vec{R}_J$ near the
junction, which is weakly coupled to the wire ($\lambda$ is small).
And we set a sufficiently
small $\epsilon$ so that the Andreev levels are close to zero energy.

\emph{Tunneling spectroscopy at the junction}.--
We use the local tunneling spectroscopy at the Josephson junction
to distinguish the topological superconducting phase from the trivial phase.
We attach a side lead (or a STM tip), which has a fixed density of states at the Fermi energy,
close to the junction.
The differential conductance in the side lead at a voltage bias $E$ is expressed as\cite{liu12,sun1,sun2}
\begin{equation}\label{conductance}
G(E) =  \Gamma _{Le} G_{eh} ^r(E) \Gamma _{Lh} G_{eh}^a(E) ,
\end{equation}
where $\Gamma_{Le}=\Gamma_{Lh}=0.1 t$ are the line-widths of the side
lead with the electron's part and the hole's part, respectively. $G^{r(a)}_{eh}$ is
the retarded (advanced) Green's function of the system (see more details of the calculation in
Ref.~\onlinecite{liu12}). We set the temperature to zero in Eq.~(\ref{conductance}).
But our conclusions keep valid at finite temperatures.

Changing the chemical potential $\mu$ tunes the topological order of the system.
We first study the conductance function $G(E)$ at different chemical potentials,
while fixing the phase difference across the junction to $\phi=\pi$ (see Fig.~\ref{fig:mu}).
In the absence of the impurity Andreev levels ($\lambda=0$ in Eq.~(\ref{manualAndreev})),
there are a unique pair of Majorana states located at the junction in the TSC phase, but no
in the trivial phase. This pair of Majorana fermions have the same symmetry, which leads to a constructive interference
in the tunneling process. Since $G(E)$ reflects the local density of states at the junction,
a ZBP emerges in $G(E)$ in the TSC phase, but is absent in the trivial phase.
In Fig.~\ref{fig:mu}(a), we clearly see a group of bright narrow lines at $E=0$
which distinguish the TSC regions from the trivial regions.
There are totally five TSC regions, corresponding to $1$, $3$, $5$, $7$ and $9$ occupied subbands respectively.

Unfortunately, a pair of impurity Andreev levels close to zero energy will cause
a similar ZBP in whatever the TSC or the trivial phases. This is shown in
Fig.~\ref{fig:mu}(b), where we choose the small but non-zero $\lambda$ and $\epsilon$.
Compared with Fig.~\ref{fig:mu}(a), a bright line at $E\approx0$ is now through the whole region of $\mu$,
so that one cannot
distinguish the TSC and the trivial phases any more.
With the model $(\hat H_w+\hat H_d)$ we successfully reproduce the results in
Refs.~\onlinecite{liu12,bagrets12}, i.e., a ZBP by itself cannot identify
the topological superconducting phase and the nontrivial Majorana fermions.

\begin{figure}
\includegraphics[width=1.0\linewidth]{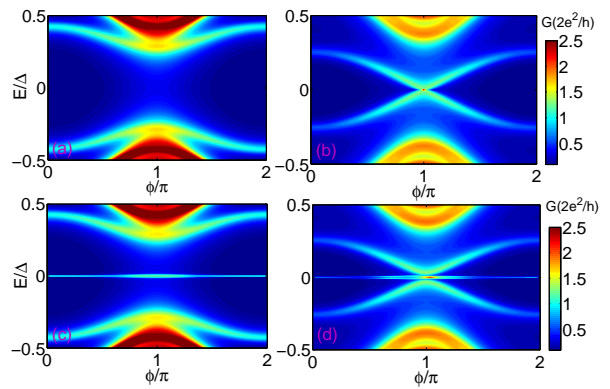}
\caption{[Color online] The evolution of the conductance as $\phi$ varies
at four different points of Fig.~\ref{fig:mu}.
(a) $\mu=-2t+4\Delta$ and $\lambda=0,\epsilon = 0.001t$. (b) $\mu=-2t$ and $\lambda=0,\epsilon = 0.001t$.
 (c)  $\mu=-2t+4\Delta$ and $\lambda=0.3, \epsilon = 0.001 t$.
(d)  $\mu=-2t$ and $\lambda=0.3,\epsilon = 0.001t$. }\label{fig:cond}
\end{figure}

We propose to distinguish the TSC phase from the trivial phase by
the behavior of the conductance as the superconducting phase difference $\phi$ varies. In the TSC phase, 
at $\phi=\pi$ the two adjacent Majorana bound states at the junction
contribute to a conductance of $4e^2/h$. As $\phi$ deviates from $\pi$, the
two Majorana modes are coupled and push each other away
from zero energy, so that the zero-bias conductance quickly decreases and
the ZBP is split into two side peaks at finite biases.
This evolution is displayed in Fig.~\ref{fig:cond}(b), where 
the conductance function $G(E,\phi)$ shows a butterfly pattern centered at $E=0$ and $\phi=\pi$.
In the topologically trivial phase (Fig.~\ref{fig:cond}(a)), however, there is no ZBP whatever 
$\phi$ is. As $\phi$ varies, the sub-gap Andreev levels avoid each other, 
so that the zero-bias conductance is always absent, indifferent to $\phi$.
Whereas the presence of impurities, disorder effect, Kondo effect, etc.,
may induce a ZBP in the trivial phase, causing that one can not distinguish the 
TSC phase from the trivial phase by the ZBP alone~\cite{liu12}.
However, this casual ZBP keeps invariant as $\phi$ varies, and no butterfly pattern is observed 
(Fig.~\ref{fig:cond}(c)). The difference between the TSC and the trivial phases is clearly 
reflected in the tunneling spectroscopy.
Therefore, by utilizing an additional degree of freedom $\phi$, we can distinguish the 
topologically trivial and nontrivial phases according to whether the ZBP at $\phi=\pi$ 
splits while $\phi$ deviates from $\pi$.

Furthermore, this criterion keeps valid in the coexistence of the casual ZBP caused 
by the disorder effect and the ZBP by the Majorana modes.
In this dirty TSC phase, the zero-bias conductance $G(E=0)$ shows a peak
at $\phi=\pi$ due to the nontrivial Majorana fermions over the background signal contributed by the impurity levels.
As $\phi$ deviates from $\pi$, the unique ZBP is split into
three peaks: two peaks are symmetric to zero, resulting in a butterfly pattern, 
and the third one keeps at $E=0$ (Fig.~\ref{fig:cond}(d)). This behavior is significantly different from that in
the topologically trivial phase as shown in Fig.~\ref{fig:cond}(a) and (c). 
In summary, with the help of the superconducting phase difference $\phi$, 
we can distinguish all the four cases: the trivial phase without the casual ZBP, 
the TSC phase without the casual ZBP, the trivial phase with the casual ZBP, and the TSC phase with the casual ZBP.
Here a butterfly pattern in the conductance function $G(E,\phi)$ is the key characteristic of the TSC phase.

\begin{figure}
\includegraphics[width=1.0\linewidth]{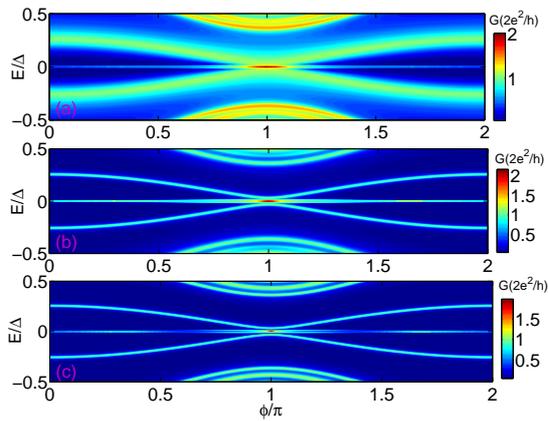}
\caption{[Color online] The 3D plot of the conductance function $G(E,\phi)$
when the side lead is coupled to different positions of the quantum wire. From top to bottom,
the distance between the side lead and the Josephson junction is (a) $1$ site, (b) $5$ sites and (c) $10$
sites, respectively. The other parameters keep the same as in Fig.~\ref{fig:cond}(d).}\label{fig:distance}
\end{figure}

The Majorana fermions at the junction are localized
in a small area of the wire, so that the side lead may not be precisely coupled to
them in experiments. We demonstrate that our conclusions are not sensitive to the
position where the side lead is coupled.
In Fig.~\ref{fig:distance}, we display the conductance $G(E,\phi)$
as the side lead gradually moves away from the junction, whereas the
impurity levels are present and fixed at the junction.
The fingerprint of the TSC phase, i.e., the butterfly pattern,
can be clearly observed even if the side lead has been
$10$ sites far away from the junction (corresponding to $200 nm$ with our parameters).
As the side lead moves away, its coupling with
the Majorana fermions becomes weaker, so that the conductance peak
becomes narrower ($\Delta \phi$ decreases). But the height of the peak always keeps invariant at zero temperature.
Whereas at finite temperatures, the conductance peak will be suppressed
when the side lead leaves the junction. But the butterfly pattern is still distinguishable,
as long as the temperatures $k_BT$ is no greater than the ZBP splitting, 
i.e., about $0.2\Delta$ in Fig.~\ref{fig:distance}.

\emph{Conclusions}.--
We conclude that the topological superconducting phase in a quantum wire
can be distinguished from the trivial phase by the behavior of the
side-lead conductance as the phase difference across the Josephson junction varies.
The TSC phase is identified by a conductance peak in $G(\phi)$ at $\phi=\pi$.
A more elaborate criterion is a butterfly pattern in the function $G(E,\phi)$, i.e., 
the splitting of the zero bias peak as $\phi$ deviates from $\pi$.
In the topologically trivial phase, however, there is either no zero-bias peak
or a casual zero-bias peak that keeps invariant as $\phi$ varies.
This offers an exclusive sign for the topological superconducting phase.

\emph{Acknowledgement}.--
This work is supported by NSFC (Grants Nos. 11304280, 
91221302 and 11274364), NBRP of China
(Grants Nos. 2012CB821402, 2012CB921303 and SQ2015CB090460) and
CPSF (Grant No. 2014M550542).

\noindent
* P. Wang and J. Liu contributed equally to this work.

\end{document}